# The diluted 2D Ising model and Non-conventional Superconductors

Xuan Zhong Ni

*Santa Rosa, California, US*

(April 2015)

Abstract: This paper demonstrates that the results of a Monte Carlo simulation of a diluted 2D Ising anti-ferromagnet system corresponds with the phase diagram for non-conventional superconductors. A new definition of energy gap in this system is presented. We also find a strange phenomenon that when the lattice size of simulation increased the crystal structure becomes more like quasi-crystal at the low temperature.

We studied the behavior of a site-diluted 2D Ising anti-ferromagnet using a Monte Carlo method.

We consider a 2D square lattice, where we have a Spin-1/2 particle on some points of the lattice. Each occupied site will either be spin up or spin down. It is important to note that these spins must be coupled with their nearest neighbors. For anti-ferromagnets, we chose the coupling so that anti-parallel spins will have less energy than parallel spins. This is achieved by

$$H = \Sigma_{\langle i,j \rangle} \; j_{\langle i,j \rangle} \; s_i s_j \qquad (1)$$

where $s_i$ denotes the spin state of the i-th site on the lattice, with $s_i = 1$ for spin up and $s_i = -1$ for spin down. $\langle ij \rangle$ denotes the sum over nearest neighbors. $j_{\langle i,j \rangle}$ is a coupling constant which should be negative one for an anti-ferromagnet to achieve the desired properties described above.

In a site-diluted Ising model, there are unoccupied sites. Thus, $j_{\langle i,j \rangle} = 0$ when one of the paired sites at $\langle ij \rangle$ is occupied and the other is unoccupied.

We also define $j_{\langle i,j \rangle} = 0$ when the two spins of the nearest-neighbor pair are parallel and $j_{\langle i,j \rangle} = -1$ when the two spins of the nearest neighbor pair are anti-parallel. This anti-parallel pair will be referred to as connected.

The Metropolis algorithm is one way to simulate the dynamics of an Ising model. For a given density of occupied sites:

1. Each randomly-chosen occupied site is assigned a value of spin up or spin down.
2. A trial configuration is made by randomly choosing a pair of lattice points and flipping the spin at one of the points and exchanging the contents of the two sites. The energy difference, deltaE, between the exchanged and the old state is then calculated.

3. If deltaE < 0, the trial state is energetically favorable and thus accepted. Otherwise, a random number η is generated and the new state is only accepted if exp(-deltaE/T) > η .

Thus, there is a probability that the system can enter higher energy states. This probability is zero at zero temperature and increases as the temperature increases. Therefore, this may be interpreted as the thermal motion of the spins.

The Monte Carlo simulation can define an energy gap for the system because of this probability.
From exp(-deltaE/T) > η, we have deltaE/T < - ln η, so the new trial configuration is accepted only when deltaE/T is under the curve of - ln η. The probability of acceptance is η.

For T < Tc, we can reasonably define $η_0$, when the line of $η = η_0$ seperates the area under the curve −ln η(0 < η < 1) to two areas. We can define $deltaE_0$ by,
   $deltaE_0/T$ = - ln $η_0$,

This $η_0$ has a special effect: after many steps we come to a stage that for any randomly choosed η, almost all of the new tried configurations with deltaE > $deltaE_0$ will be rejected if η > $η_0$. And this is exactly in the sense of Quantum Mechanics that there is a defined energy gap here.
For example we consider T at $T_c/2$ , and $η_0$ equal 0.03. This means only
3 percent of new trial configuation with deltaE > $deltaE_0$ will be energetically accepted to jump into a higher energy state.

We calculate the −ln $η_0$ = 3.507 when $η_0$ = 0.03.
And we have $deltaE_0/T_c$ = −ln $η_0$ $(T/T_c)$ = 1.75
This $deltaE_0$ is the energy gap $\Delta(T)$ and we have the following result, $\Delta(T)/T_c$ = 1.75, which is very closed to the BCS result of $\Delta(0)/T_c$ = 1.76.

The Monte Carlo simulation results using a 40x40 square lattice are shown in Table 1. The first row show the temperatures from 0.2 to 3.0, while rows 2 through 8 show the variable densities of 16/16, 15/16, 12/16, 9/16, 6/16, 3/16 and 1/16. 15 million samples were used to reach the equilibrium point, and 10 millions samples were used to obtain the averages at each data point. Here we adopt the periodical boundary condition.

The corresponding curves are shown in Figure 1.

From Fig. 1 we can see the average energy at density 16/16 approaches -1 at high temperatures and -2 at low temperatures. The maximum energy for one spin site is -2. At high temperatures, half of the nearest-neighbor pairs will be parallel ($j_{<i,j>}$ = 0) while the other half is anti-parallel ($j_{<i,j>}$ = -1). For the same reason, at density 3/16, the average energy will approach –3/16, or -0.19 at high temperatures. Therefore, at any given density x, the average energy is expected to approach –x at high temperatures.

This also means that at high temperatures, the ratio of connected pairs to the total number of nearest-neighbor pairs is x/2.

From the curves, we can locate the critical points of phase transitions for the different densities. As shown in Table 2, the $T_c$ will be 0.35, 0.46, 0.54, 0.68, 0.82, 1.10, 1.27 for density 1/16, 3/16, 6/16, 9/16, 12/16, 15/16 and 16/16 respectively.

As the occupied site density increases, the critical temperature increases. A smaller density implies that there are fewer connected pairs. Thus, it is easier for thermal fluctuations to overcome the tendency towards long-range order. In the ideal case the critical temperature becomes zero when the density reaches zero.

In this model each connected pair contributes -1 to the average energy calculation. When we choose the two sites randomly and flip the spin at one of the two sites (if it has a spin), the number of connected pairs broken could be 4, 3, 2, or 1. It is obvious that for a heavily diluted case (very low density), the probability of breaking 4 connected pairs during this exchange would be higher than at higher densities. Thus, as the density increases, the number of connected pairs being broken decreases.

We can define a Pseudo Temperature, $T^*$, as $\exp(-(m/T^*)) = ½$, or $T^* = m/(\ln(2))$.

The number of connected pairs being broken, m, is determined by the density of occupied sites.

At very low density we will have m = 4 and $T^* = 5.76$. At higher densities, such as x = 16/16 = 1, we have m = 1 and $T^* = 1.44$. This is close to the critical temperature of 1.27 as from the simulation. The results are shown in Table 2 and Figure 2.

For temperature $T > T^*$, the thermal fluctuations will overcome the tendency towards long-range order. For lower temperatures $T < T^*$ at any given density, the system will form a domain with long-range order.

The plotted diagram in Figure 2 corresponds to the phase diagram of nonconventional superconductors with an optimal doping ratio $\delta_o = 1/n$, where n corresponds to the number of the layers needed to form a super-cell of the 2D lattice. For example, in $YBa_2Cu_3O_7$ ($\delta_o = ½$), two $CuO_2$ layers form one super-cell lattice. At the optimal doping ratio, each super-cell only has one itinerant electron.

For a $La_{2-x}Sr_xCuO_4$ sample, the optimal doping ratio is x = 0.2. Therefore, it needs five layers of $CuO_2$ to form a super-cell lattice in order to be reach density one for an optimal $T_c$ sample.

Table 1  Monte Carlo Simulation of Diluted 2D Ising Model

of variable density of 16/16, 15/16, 12/16, 9/16, 6/16, 3/16 and 1/16 from row 2 to row 8, the 1st raw is the corresponding temperatures:

| 0.2  | 0.4  | 0.6  | 0.8  | 1    | 1.2  | 1.4  | 1.6  | 1.8  | 2    | 2.2  | 2.4  | 2.6  | 2.8  | 3    |
|------|------|------|------|------|------|------|------|------|------|------|------|------|------|------|
| -2   | -2   | -2   | -2   | -2   | -1.7 | -1.3 | -1.3 | -1.2 | -1.2 | -1.1 | -1.1 | -1.1 | -1.1 | -1.1 |
| -2   | -2   | -2   | -2   | -1.8 | -1.3 | -1.2 | -1.2 | -1.2 | -1.1 | -1.1 | -1.1 | -1   | -1   | -1   |
| -2   | -2   | -1.8 | -1.8 | -1.2 | -1.1 | -1   | -1   | -1   | -1   | -1   | -1   | -1   | -1   | -1   |
| -2   | -2   | -1.7 | -1.1 | -1   | -1   | -0.9 | -0.9 | -0.8 | -0.7 | -0.7 | -0.6 | -0.6 | -0.6 | -0.6 |
| -2   | -1.9 | -1.2 | -1   | -0.6 | -0.6 | -0.5 | -0.5 | -0.5 | -0.5 | -0.5 | -0.5 | -0.5 | -0.5 | -0.4 |
| -2   | -1.8 | -0.6 | -0.5 | -0.4 | -0.3 | -0.3 | -0.3 | -0.3 | -0.3 | -0.3 | -0.3 | -0.3 | -0.3 | -0.3 |
| -1.9 | -1.2 | -0.3 | -0.2 | -0.2 | -0.1 | -0.1 | -0.1 | -0.1 | -0.1 | -0.1 | -0.1 | -0.1 | -0.1 | -0.1 |
|      |      |      |      |      |      |      |      |      |      |      |      |      |      |      |

Table 2  The critical temperature Tc and pseudo temperatures T*

| X  | 1/16 | 3/16 | 6/16 | 9/16 | 12/16 | 15/16 | 16/16 |
|----|------|------|------|------|-------|-------|-------|
| Tc | 0.35 | 0.46 | 0.54 | 0.68 | 0.82  | 1.10  | 1.27  |
| T* | 5.76 |      | 4.32 |      | 2.88  |       | 1.44  |

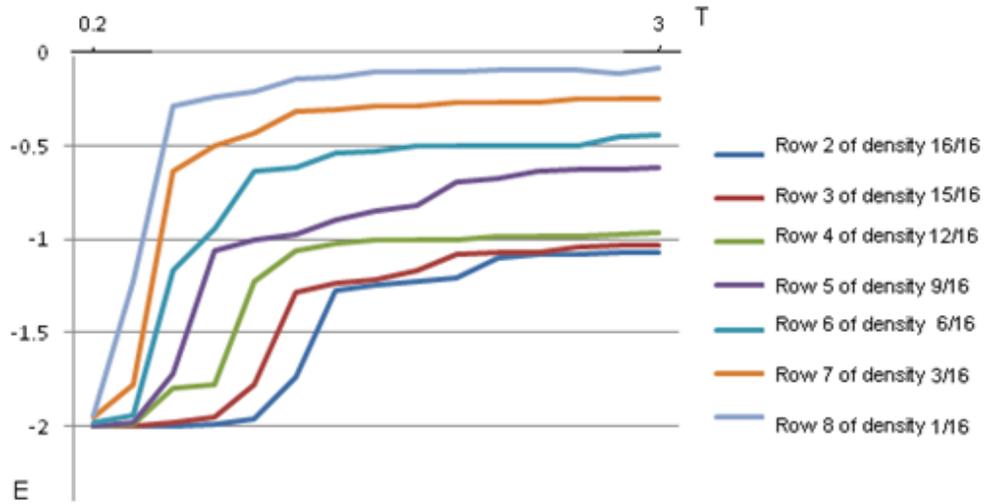

Figure 1, Curves are from the data in Table 1 of variable density verses the Temperature

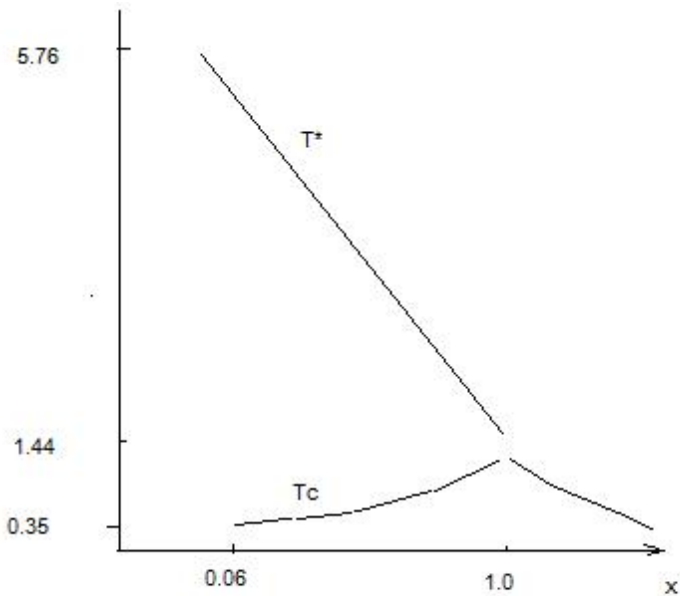

Figure 2
Phase Diagram

In our previous paper[1], we described the non-conventional superconductors as 2d-lattice of super-cells with the $p_z$ itinerant electrons or holes in a cloud layer over the super-cell physical layers with the Hamiltonian as follows:

$$H = U/3 \sum_{<i,j> \sigma} \{ (\pi_{ij\sigma} \pi_{ij\sigma'}) \} \qquad (2)$$

with $\pi_{ij\sigma} = C^+_{j\sigma} C_{i\sigma} + C^+_{i,\sigma} C_{j\sigma}$ and σ' having the opposite spin of σ.

We have,

$$(\pi_{ij\sigma})^2 = (C^+_{j\sigma} C_{i\sigma} + C^+_{i,\sigma} C_{j\sigma})^2 = n_{i\sigma} + n_{j\sigma} - 2n_{i\sigma}n_{j\sigma}$$

with $n_{i\sigma} = C^+_{i,\sigma} C_{i\sigma}$.

$(\pi_{ij\sigma})^2$ equals zero when $n_{i\sigma}$ and $n_{j\sigma}$ both equal 0 or 1, and it will equal 1 when only one of $n_{i\sigma}$ or $n_{j\sigma}$ equals one.

In the $\{n_{i\sigma}\}$ space, the $\pi_{ij\sigma}$ are off diagonal Hermitian operators with eigenvalues of plus and minus one.

If we take out U/3 to remove dimensional parameters, the Hamiltonian (2) will give rise the same results as Hamiltonian (1) since the stable state will have the lower energy.

If M is the total number of the lattice sites and N is the total number of particles in the lattice, x = N/M is the particle density in the lattice space. If x ≤ 1, we assume each lattice site will have only one particle as in our Monte Carlo simulation with results in Table 1. From these results we obtain a phase diagram that is similar to the phase diagram of non-conventional superconductors.

When x > 1, our current model would have to be adjusted. In our modeling of (2), $(\pi_{ij\sigma})^2$ will become zero more frequently as x gets larger. If we start with itinerant electrons at x < 1, then when x > 1 we must account for itinerant holes of density (2 – x). Of course, the Pauli Principle prevents x from having a value greater than 2.

The Hall coefficient will be follows;

$$H = 1/(xe) - 1/((2-x)e) = 2 (1 - x)/[ x (2 - x)e] \qquad (3)$$

The doping ratio of itinerant electrons and holes explains the anomalous Hall effects[2] of sign changes in non-conventional superconductors. At optimal doping (x=1), the total number of itinerant electrons equals the total amount of itinerant of holes, so the net itinerant charge is neutral. This explains why the Hall coefficient is very small or zero. When x is small the Hall coefficient is approximately 1/(xe), and when x is close to 2 the Hall coefficient is closed to - 1/((2-x)e) with the sign changed.

Given the symmetry of itinerant electron and holes, we should expect a symmetric T* curve on the opposite side when 2 > x > 1. This will be a subsequent test for our theory.

Table 1 shows the Monte Carlo simulation results using a 40x40 square lattice with periodic boundary. We also did simulation with 100x100, 200x200, 300x300 and 400x400 square lattices. We were able to achieve similar results as Table 1 and similar critical temperatures Tc for different densities. But we have found a strange phenomenon that while the lattice size becomes larger, more small domains appears until the domain ceased to aggregate together at lattice 400x400, at the temperatures equal 0.4 which is lower than the Tc. The final spin array outputs are shown in Fig. 3 to Fig. 6. Fig. 6 is the result with Monte Carlo simulation by one trillion steps.

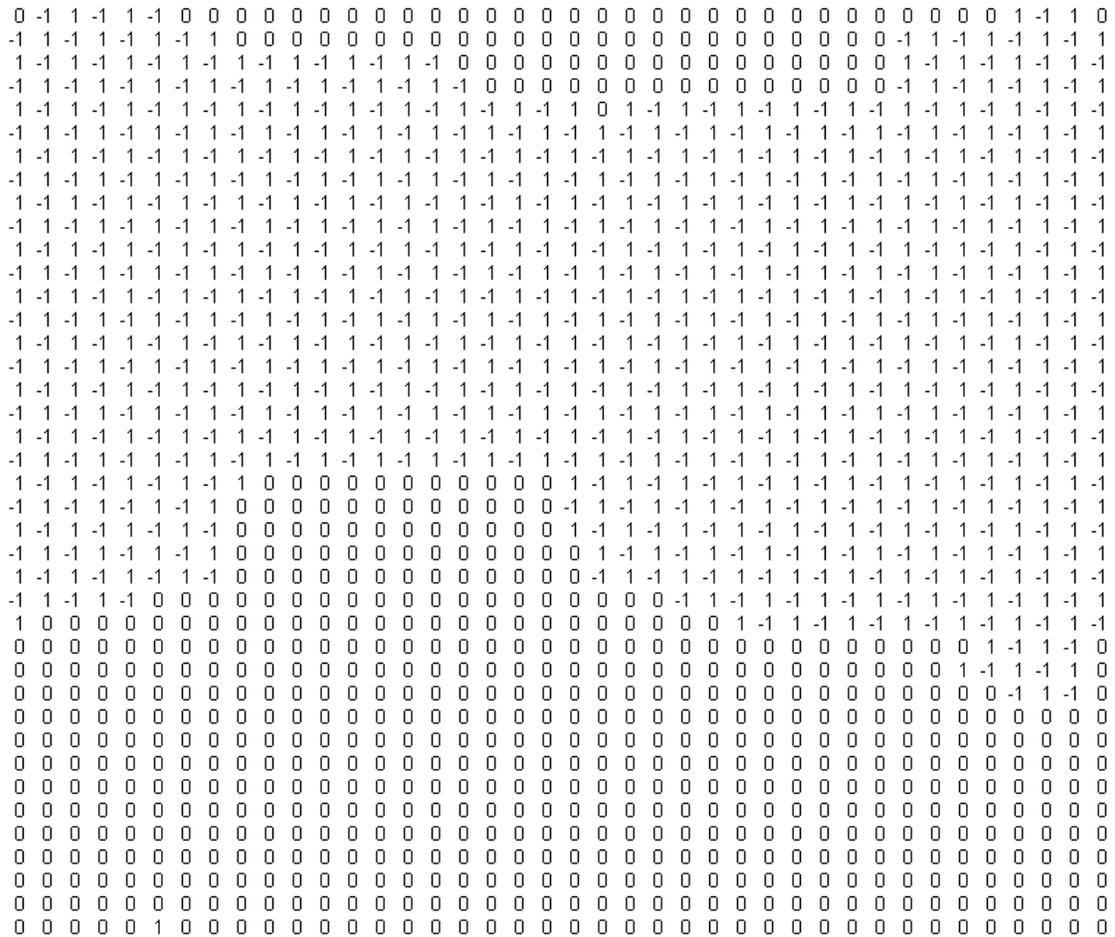

Fig. 3  40x40 lattice with density 9/16 at T=0.4, here zero for empty sites, plus and minus one are for sites with spin up and down,  and obviously the domain aggregated together.

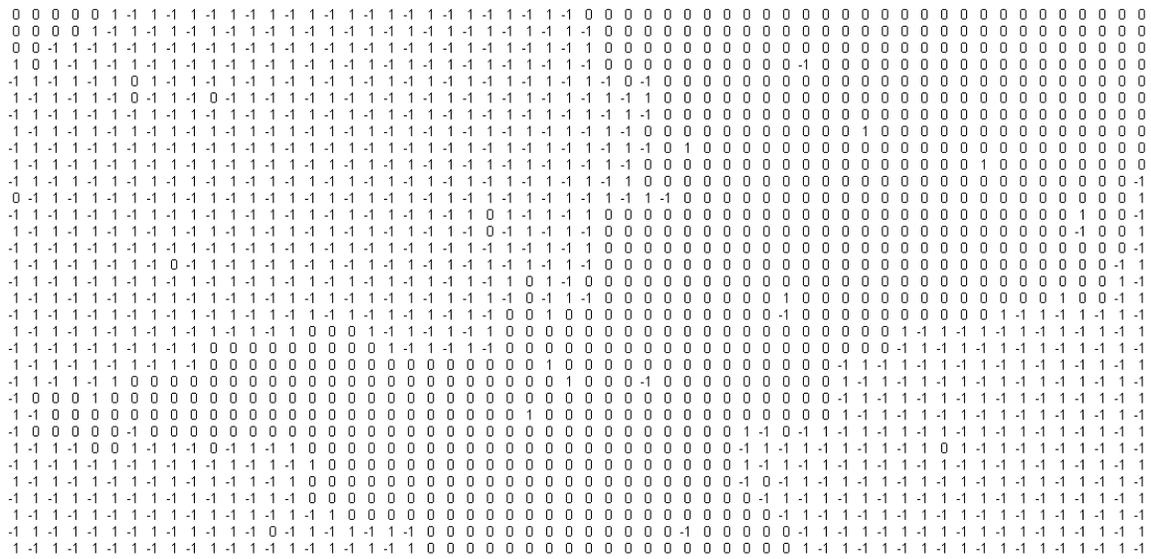

Fig. 4, part of 100x100 lattice with density 9/16 at T=0.4, here zero for empty sites, plus and minus one are for sites with spin up and down, and obviously the domain is still aggregated together but some small isolated domains appear.

```
 1  0  0  1 -1  0  0  0  0  1 -1  1 -1  1 -1  1 -1  1 -1  0  0  0 -1  1 -1  1 -1  1 -1  1 -1  1 -1  1  0  1  0  0  0
-1  0  0 -1  0  0  0  0  0 -1  1 -1  1 -1  1 -1  1 -1  1 -1  1 -1  1 -1  1 -1  1 -1  1 -1  1 -1  1 -1  1 -1  0 -1  0  0  0
 1  0  0  0  0  1  0  0  1 -1  1 -1  1 -1  1 -1  1 -1  1 -1  1 -1  1 -1  1 -1  1 -1  1 -1  1 -1  1 -1  1 -1  1 -1  1 -1
 0  0  0  0  0  0  0  0 -1  1 -1  1 -1  1 -1  1 -1  1 -1  1 -1  1 -1  1 -1  1 -1  1 -1  1 -1  1 -1  1 -1  1 -1  1 -1  1
 0  0  0  0  0  0  0  0  0  1 -1  1 -1  1 -1  1 -1  1 -1  1 -1  1 -1  1 -1  1 -1  1 -1  1 -1  1 -1  1 -1  1 -1  1 -1
 0  0  0  0  0  0  0  0  0 -1  1 -1  1 -1  1 -1  1 -1  1 -1  1 -1  1 -1  1 -1  1 -1  1 -1  1 -1  1 -1  1 -1  1 -1  1
 0  0  0  0  0  0  0  0  0  1 -1  1 -1  1 -1  1 -1  1 -1  1 -1  1 -1  1 -1  1 -1  1 -1  1 -1  1 -1  1 -1  1 -1  1 -1
 0  0  0  0  0  0  0  1  0  1 -1  1 -1  1 -1  1 -1  1 -1  1 -1  1 -1  1 -1  1 -1  1 -1  1 -1  1 -1  1 -1  1 -1  1 -1  1
 0  0  0  0  0  0 -1  1 -1  1 -1  1 -1  1 -1  1 -1  1 -1  1 -1  1 -1  1 -1  1 -1  1 -1  1 -1  1 -1  1 -1  1 -1  1 -1  1
 0  0  0  0  0  0  1 -1  1 -1  1 -1  1 -1  1 -1  1 -1  1 -1  1 -1  1 -1  1 -1  1 -1  1 -1  1 -1  1 -1  1 -1  1 -1  1 -1
-1  0  0  0  0 -1  1 -1  1 -1  1 -1  1 -1  1 -1  1 -1  1 -1  1 -1  1 -1  1 -1  1 -1  1 -1  1 -1  1 -1  1 -1  1 -1  1 -1
 0  0  0  0  0  0 -1  1 -1  1 -1  1 -1  1 -1  1 -1  1 -1  1 -1  1 -1  1 -1  1 -1  1 -1  1 -1  1 -1  1 -1  1 -1  1 -1  1
 0  0  0  0  0  0  0 -1  1 -1  1 -1  1 -1  1 -1  1 -1  1 -1  1 -1  1 -1  1 -1  1 -1  1 -1  1 -1  1 -1  1 -1  1 -1  1 -1
 0  0  0  0  0  0  0 -1  1 -1  1 -1  1 -1  1 -1  1 -1  1 -1  1 -1  1 -1  1 -1  1 -1  1 -1  1 -1  1 -1  1 -1  1 -1  1 -1
 0  0  0  0  0  0  0  0  1 -1  1 -1  1 -1  1 -1  1 -1  1 -1  1 -1  1 -1  1 -1  1 -1  1 -1  1 -1  1 -1  1 -1  1 -1  1 -1
 0  0  0  0  0  0  0  1 -1  1 -1  1 -1  1 -1  1 -1  1 -1  1 -1  1 -1  1 -1  1 -1  1 -1  1 -1  1 -1  1 -1  1 -1  1 -1  1
 0  0  0  0  0  0  0 -1  1 -1  1 -1  1 -1  1 -1  1 -1  1 -1  1  0  0  0  0 -1  1 -1  1 -1  1 -1  1 -1  1 -1  1 -1  1 -1
 0  0  0 -1  1 -1  1 -1  1 -1  1 -1  1 -1  1 -1  1 -1  1  0  0  0  0  0  0  0  0  1 -1  1 -1  1 -1  1  0 -1  1 -1  1
 0  0  0  0  0  1 -1  1 -1  1 -1  1 -1  1 -1  0 -1  0  0  0  0  0  0  0  0  0  0 -1  1 -1  1  0 -1  1 -1  1 -1  1  0
 0  0  0  0  0 -1  1 -1  1 -1  1 -1  1 -1  1  0  0  0  0  0  0  0  0  0  0  0  0  0 -1  1 -1  1 -1  1 -1  1 -1  1 -1  0
 0  0  0  0  0  1  0  1 -1  1 -1  1 -1  1 -1  0  0  0  0  0  0  0  0  0  0  0  0  0  0 -1  1 -1  1 -1  1 -1  1 -1  1 -1
 0  0  0  0  0 -1  1 -1  1 -1  1 -1  1 -1  1  0  0  0  0  0  0  0  0  0  0  0  0  0  1 -1  1 -1  1 -1  1 -1  1 -1  1
 0  0  0  0  0  0 -1  1 -1  1 -1  1 -1  1  0  0  0  0  0  0  0  0  0  0  0  0  0  0 -1  1 -1  1 -1  1 -1  1 -1  1  0
 0  0  0  0  0  0  1 -1  1 -1  1 -1  1 -1  1  0  0  0  0  0  0  0  0  0  0  0  0  0  0  0  0  1 -1  1  0  0  0  0  0
 0  0  0  0  0  0 -1  1 -1  1 -1  1 -1  1  0  0  0  0  0  1 -1  0  0  0  0  0  0  0  0  0  0  0  0  0  0  0  0  0  0
 0  0  0  0  0  0  1 -1  1 -1  1 -1  1 -1  0  0  0  0  0  0  0  0  0  0  1  0  0  0  0  0  0  0  0  0  0  0  0  0  0
 0  0  0  0  0 -1  0 -1  1 -1  1 -1  1 -1  1  0  0  0  0  0  0  0  0  0 -1  1  0  0  0  0  0  0  0  0  0  0  0  0  0  0
 0  0  0  0  0  0  1  0  1 -1  1 -1  1  0  0  0  0  0  0  0  0  0  0  0  1 -1  0  0  0  0  0  0  0  0  0  0  0  0  0  0
 0  0  0  0  0  0  0 -1  1 -1  1  0  0  0  0  0  0  0  0 -1  1 -1  1 -1  1 -1  1  0  0  0  0  0  0  0  0  0  0  0  0  0
 0  0  0  0  0  0  0  1 -1  1  0  0  0  0  0  0  0  0  0  1 -1  1 -1  1 -1  1 -1  1  0  0  0  0  0  0  0  0  0  0  0  0
 0  0  0  0  0  0  0  0  0  0  0  0  0  0  0  0  0  1 -1  1 -1  1 -1  1 -1  1 -1  1 -1  0  0  0  0  0  0  0  0  0  0  0
 0  0  0  0  0  0  0  0  0  0  0  0  0  0  0  1 -1  1 -1  1 -1  1 -1  1 -1  1 -1  1 -1  0  0  0  0  0  0  0  0  0  0  0
 0  0  0  0  0  0  0  0  0  0  0  0  0  0 -1  1 -1  1 -1  1 -1  1 -1  1 -1  1 -1  1 -1  1  0  0  0  0  0  0  0  0  0  0
 0  0  0  0  0  0  0  0  0  0  0  0  0  0  1 -1  1 -1  1 -1  1 -1  1 -1  1 -1  1 -1  1  0  0  0  0  0  0  0  0  0  0  0
 0 -1  1 -1  0 -1  1  0  0  0  0  0  0  0  1 -1  1 -1  1 -1  1 -1  1 -1  1 -1  1 -1  1  0  0  0  0  0  0  0  0  0  0  0
-1  1 -1  1 -1  1 -1  1  0  0  0  0  0  0 -1  1 -1  1 -1  1 -1  1 -1  1 -1  1 -1  1 -1  1  0  0  0  0  0  0  0  0  0  0  0
 1 -1  1 -1  1 -1  1 -1  1  0  0  0  0  0  1 -1  1 -1  1 -1  1 -1  1 -1  1 -1  1 -1  1 -1  0  0  0  0  0  0  0  0  0  0  0
-1  1 -1  1 -1  1 -1  1 -1  1  0  0  0  0 -1  1 -1  1 -1  1 -1  1 -1  1 -1  1 -1  1 -1  1  0  0  0  0  0  0  0  0  0  0  0
 1 -1  1 -1  1 -1  1 -1  1 -1  0  0  0  0  0 -1  1 -1  1 -1  1 -1  1 -1  1 -1  1 -1  1 -1  0  0  0  0  0  0  0  0  0  0  0
-1  1 -1  1 -1  1 -1  1 -1  1 -1  1  0  0  0  0  0  0  0 -1  1 -1  0  0  0  0  0  0  0  0  0  0  0  0  0  0  0  0  0  0
 1 -1  1 -1  1 -1  1 -1  1 -1  0  0  0  0  0  0  0  0  0  0  0  0  0  0  0  0  0  0  0  0  0  0  0  0  0  0  0  0  0  0
```

Fig. 5, part of 200x200 lattice with density 9/16 at T=0.4, the domain is still aggregated together but more small isolated domains appear.

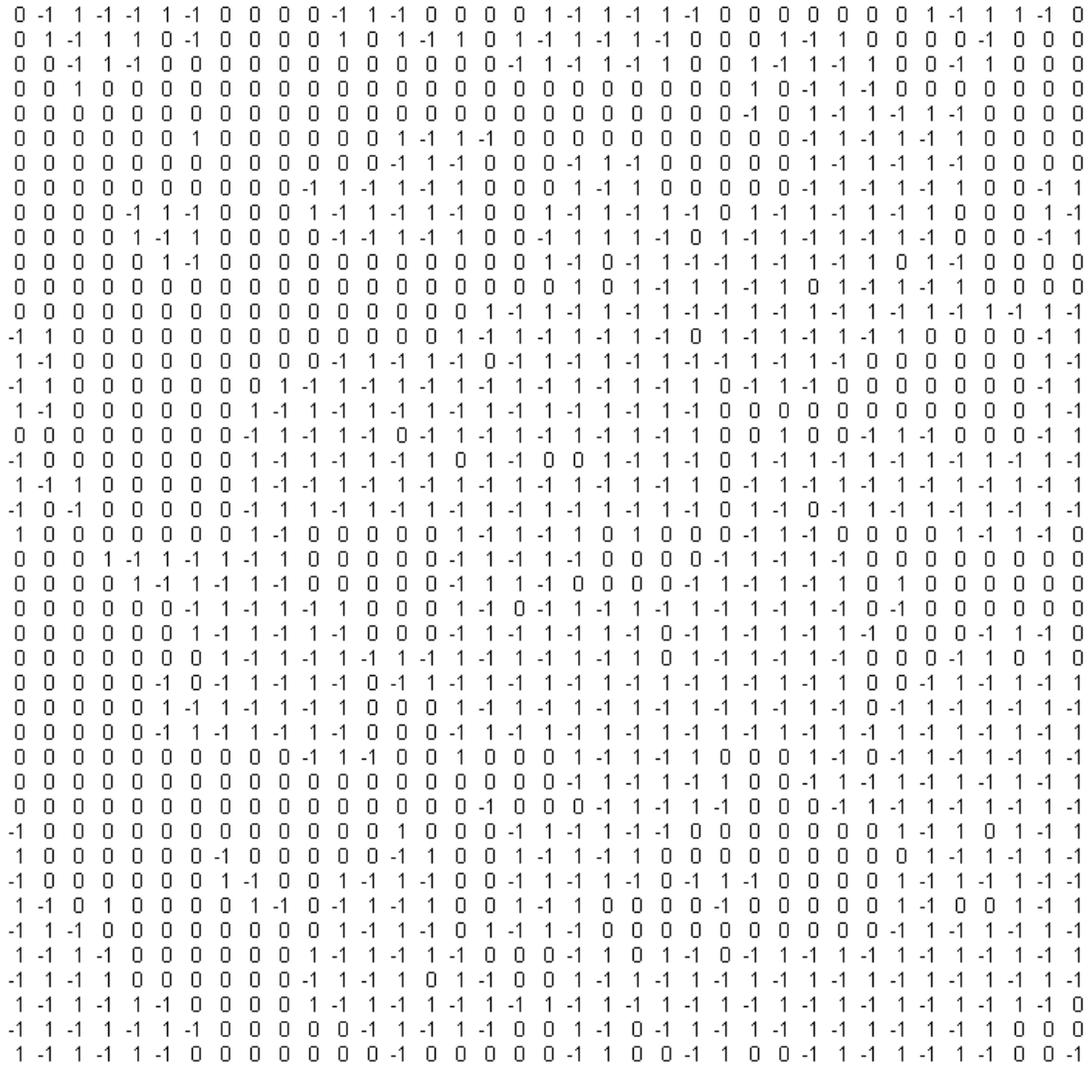

Fig. 6, part of 400x400 lattice with density 9/16 at T=0.4, the domain is no longer aggregated all together while more small isolated domains appear and the periodic structure is more like a quasi-crystal.

Reference;